\documentclass[showpacs,10pt,twocolumn,prb]{revtex4}
\usepackage{amssymb}
\usepackage{graphics}
\usepackage{epsfig}
\usepackage{graphicx}

\graphicspath{{converted_graphics/}}
\begin{document}

\title{Magnetism, Superconductivity and Stoichiometry in Single Crystals of
Fe$_{1+y}$(Te$_{1-x}$S$_{x}$)$_{z}$ }
\author{Rongwei Hu$^{1}$, Emil S. Bozin$^{1,2}$ and C. Petrovic$^{1}$}
\affiliation{$^{1}$Condensed Matter Physics and Materials Science Department, Brookhaven
National Laboratory, Upton, NY 11973}
\affiliation{$^{2}$Department of Applied Physics and Applied Mathematics, Columbia
University, New York, New York, 10027, USA}
\date{\today }

\begin{abstract}
We report synthesis of high quality Fe$_{1+y}$(Te$_{1-x}$S$_{x}$)$_{z}$
single crystals and a comprehensive study of structural, magnetic and
transport properties. There is high sensitivity to material stoichiometry
which includes vacancies on the Te(S) site. Our results reveal competition
and coexistence of magnetic order and percolative superconductivity for x $%
\geq $ 0.03, while zero resistivity is acheived for x $\geq $ 0.1.
\end{abstract}

\maketitle

\section{Introduction}

All exotic superconductors exhibit Cooper pairing in proximity to a magnetic
ground state, regardless of crystal structure and bonding type: cuprate
oxides, heavy fermion intermetallics and organics. Recently discovered iron
pnictide superconductors are no exception: magnetism in these materials is
strongly influenced by subtle crystal structure changes.\cite{Hosono}$^{-}$%
\cite{Mike} Binary iron chalcogenides FeSe and FeTe share square planar
layers of tetrahedrally coordinated Fe with the Fe-As based superconductors,
yet they crystallize in a simple crystal structure which is amenable to
modeling by band structure calculations.\cite{Hsu}$^{-}$\cite{Han} FeSe
hosts high $T_{c}$'s of up to 37 K under pressure and an isotropic
superconducting state. Its crystal structure changes from high temperature
tetragonal \textit{P4/nmm} to low temperature orthorhombic \textit{Cmmm} at
70 K.\cite{Margadonna}$^{-}$\cite{Margadonna2} In Fe$_{1.08}$Te transition
to monoclinic space group \textit{P2}$_{1}$\textit{/m} with commensurate AF
order occurs between 65 and 75 K, whereas Fe$_{1.14}$Te exhibits weaker
first-order transition to orthorhombic space group \textit{Pmmm} and
incommensurate AF order from 55 K to 63 K.\cite{Bao} In this work we examine
the evolution of superconductivity and magnetism in single crystals of Fe$%
_{1+y}$(Te$_{1-x}$S$_{x}$)$_{z}$ for $x=(0-0.15)$, $y=(0-0.14$) and $%
z=(0.94-1)$. We provide experimental evidence for structural parameters at
the magnetic/superconducting boundary.

\section{Experimental Method}

Single crystals were grown from Te(S) flux and their composition was
determined as previously described.\cite{Rongwei} Total scattering data from
finely pulverized crystals were obtained at 80 K at 11-ID-B beamline of the
Advanced Photon Source synchrotron using 58.26 keV x-rays ($\lambda $ =
0.2128 \AA ) selected by a Si 311 monochromator. 2D patterns for samples in
1mm diameter Kapton tubes were collected using a MAR345 2D-detector, placed
perpendicular to the primary beam path, 188.592 mm away from the sample. An
Oxford Cryosystem cryostream was used for temperature regulation. Details on
experimental procedures, data processing, the atomic pair distribution
function (PDF) method, and structural modeling can be found elsewhere.\cite%
{Chupas}$^{-}$\cite{Farrow} Electrical resistivity measurements were
performed using a standard four probe method with current flowing parallel
to the $\widehat{a}$ - axis of tetragonal structure. Magnetization and
resistivity measurements were carried out in a Quantum Design MPMS-5 and a
PPMS-9 instruments.

\begin{figure}[tbh]
\centerline{\includegraphics[scale=0.3]{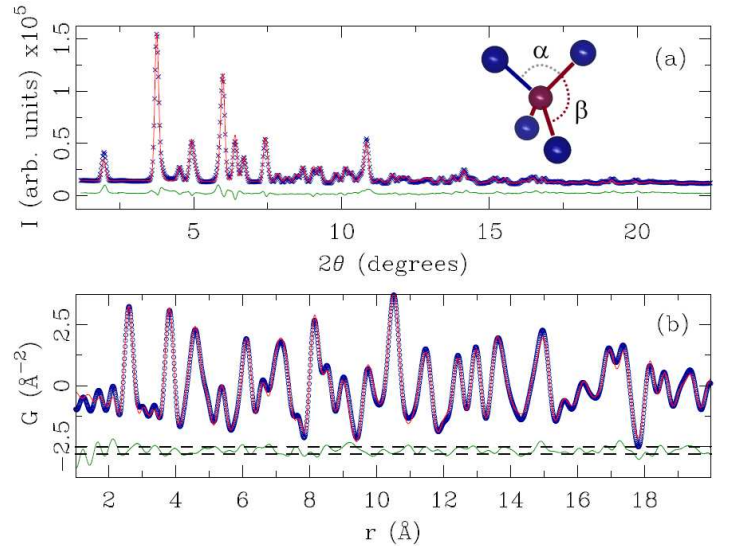}} 
\vspace*{-0.3cm}
\caption{Fe$_{1+y}$(Te$_{1-x}$S$_{x}$)$_{z}$ synchrotron Rietveld (a) and
PDF refinement (b) results taken at T = 80 K. Inset shows two Te-Fe-Te bond
angles illustrated on the FeTe$_{4}$ tetrahedron.}
\end{figure}

\begin{table*}[tph]
\caption{{\protect\small Structural parameters from PDF\ refinement at T =
80 K, magnetic and superconducting properties of Fe}$_{{\protect\small 1+y}}$%
({\protect\small Te}$_{{\protect\small 1-x}}${\protect\small S}$_{%
{\protect\small x}}$)$_{z}${\protect\small . Transition temperatures }$%
{\protect\small T}_{{\protect\small 1}}${\protect\small \ and }$%
{\protect\small T}_{{\protect\small 2}}${\protect\small \ are from }$%
\partial (\protect\chi T/\partial T)$ {\protect\small data. Temperatures of }%
${\protect\small T}_{{\protect\small c}}${\protect\small \ onset and zero
resistance are from resistivity data.}}%
\begin{tabular}{cccccccccccc}
\hline\hline
Fe$_{1+y}$(Te$_{1-x}$S$_{x}$)$_{z}$ & $V(\mathring{A}^{3})$ & $c/a$ & $%
Occ(Fe1)$ & $(OccFe2)$ & $\alpha $ & $\Theta (K)$ & $\mu (\mu _{B})$ & $%
T_{2}(K)$ & $T_{1}(K)$ & $T_{C}^{onset}(K)$ & $T_{C}(K)$ \\ \hline
Fe$_{1.14(2)}$Te$_{1.01(1)}$ & $91.150(4)$ & $1.642(1)$ & $1.04(2)$ & $%
0.10(2)$ & $117.46(1)$ & $-191(4)$ & $3.92(2)$ & $59(1)$ & $70(1)$ &  &  \\ 
Fe$_{1.09(2)}$Te$_{1.00(1)}$ & $91.017(4)$ & $1.640(1)$ & $1.02(2)$ & $%
0.07(2)$ &  &  & $3.73(1)$ & $59(1)$ & $66(1)$ &  &  \\ 
Fe$_{1.12(3)}$Te$_{0.97(1)}$S$_{0.03(2)}$ & $90.558(4)$ & $1.637(1)$ & $%
1.00(3)$ & $0.12(3)$ & $117.57(1)$ & $-175(1)$ & $\,3.83(1)$ & $41(1)$ & $%
44(1)$ & $6.5(1)$ &  \\ 
Fe$_{1.13(3)}$Te$_{0.85(1)}$S$_{0.10(2)}$ & $90.032(5)$ & $1.632(1)$ & $%
1.00(4)$ & $0.13(4)$ & $117.33(1)$ & $-186(4)$ & $3.56(2)$ & $21(1)$ & $%
23(1) $ & $8.5(1)$ & $2.0(1)$ \\ 
Fe$_{1.12(3)}$Te$_{0.83(1)}$S$_{0.11(2)}$ & $90.095(4)$ & $1.632(1)$ & $%
1.06(4)$ & $0.07(4)$ & $117.30(1)$ & $-162(3)$ & $3.38(2)$ &  & $20(1)$ & $%
8.6(1)$ & $3.5(1)$ \\ 
Fe$_{1.06(3)}$Te$_{0.88(1)}$S$_{0.14(2)}$ & $90.097(4)$ & $1.632(1)$ & $%
0.95(4)$ & $0.11(4)$ & $117.32(1)$ & $-156(6)$ & $3.36(3)$ &  & $23.5(1)$ & $%
8.7(1)$ & $7.0(1)$ \\ 
Fe$_{0.98(4)}$Te$_{0.90(1)}$S$_{0.15(2)}$ & $89.900(5)$ & $1.632(1)$ & $%
0.82(4)$ & $0.16(4)$ & $117.21(1)$ & $-167(7)$ & $3.34(6)$ &  & $19(1)$ & $%
8.8(5)^{\ast }$ &  \\ \hline\hline
\end{tabular}%
\end{table*}

\section{Results}

Typical synchrotron data (symbols), with fully converged \textit{P4/nmm}
structural model superimposed (solid lines), and corresponding difference
curves (offset for clarity) are presented in Figure 1, featuring Fe$_{1.12}$%
Te$_{0.83}$S$_{0.11}$. Panel (a) features a Rietveld refinement, with a
corresponding PDF refinement shown in (b). The PDF is peaked at positions
corresponding to the observed interatomic distances. Excellent fits with low
agreement factors are obtained for all studied Fe$_{1+y}$(Te$_{1-x}$S$_{x}$)$%
_{z}$ samples. Initial PDF fits assumed ideal stoichiometry, and revealed
slightly enhanced atomic displacement parameters (ADPs), suggesting either
Te(S) deficiency, or excess Fe content, or both. To explore this, in final 
\textit{P4/nmm} models where S shared site with Te, Fe/Te/S ratios were kept
fixed at values obtained from SEM, while stoichiometry was refined.
Additionally, Fe was allowed to occupy two sites, Fe(1) (0,0,0) and Fe(2)
(0.5,0,z). Total Fe content in the refinements was constrained to respect
the SEM ratio, while the relative occupancy of the two Fe sites was allowed
to vary. Results are reported in Table I. Undoped FeTe crystallizes with
excess Fe variably occupying interstitial Fe(2) site and full occupancies of
Fe(1) and Te sites. With contraction of the unit cell due to sulfur doping
we observe both excess Fe and vacancies on Te(S) site. With reduction of the
unit cell due to an increase in sulfur stoichiometry ($x$) excess Fe ($y$)
decreases so that for highest $x=0.15$ we observe the smallest deviation
from ideal stoichiometry. Both Fe(1) and Fe(2) sites are still occupied \
for high $x$ values (Table I).

\begin{figure}[tbh]
\centerline{\includegraphics[scale=0.15]{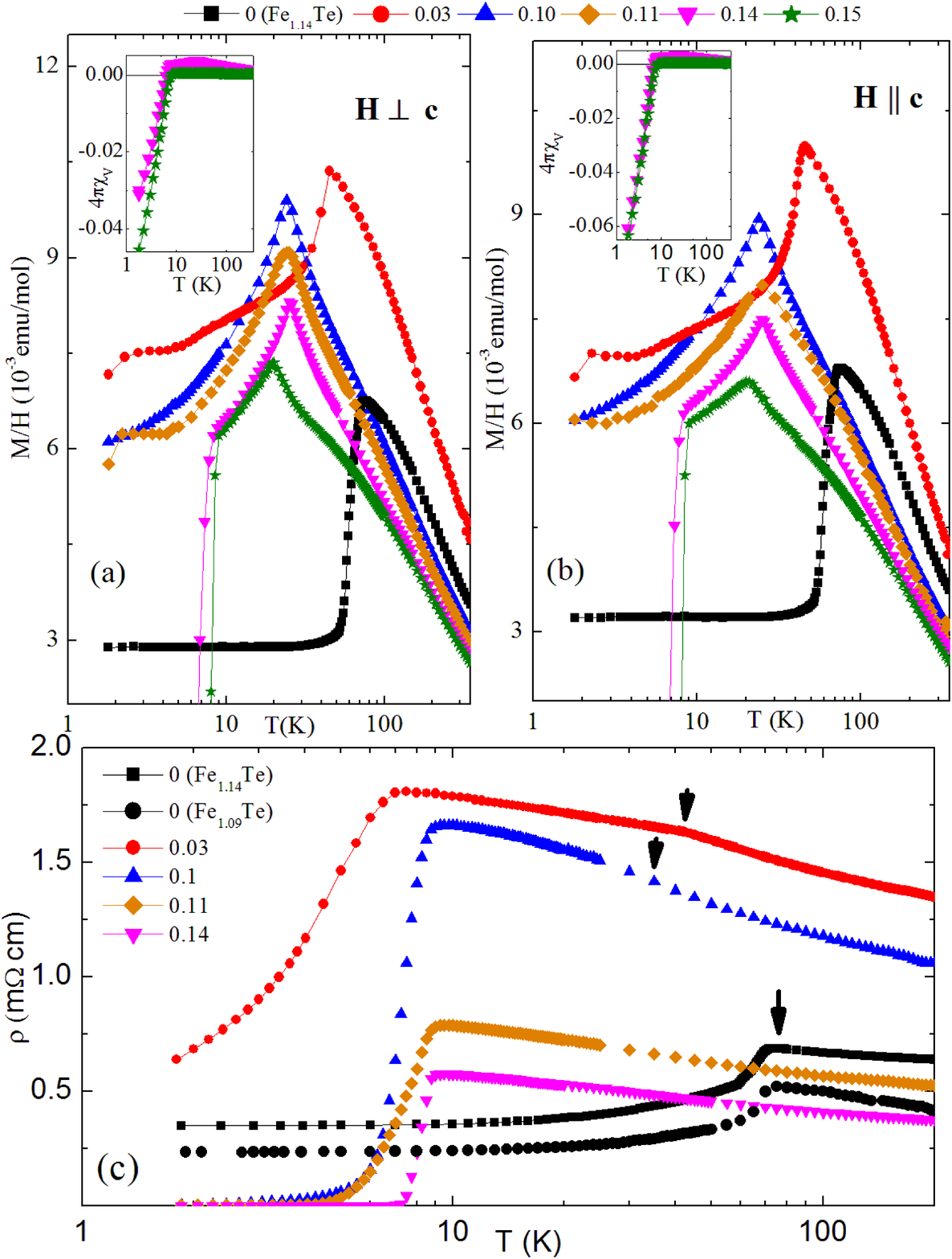}} 
\vspace*{-0.3cm}
\caption{Magnetic susceptibility as a function of temperature for $H$ $\perp
c$ (a) and $H\parallel c$ (b). Insets show the Meissner volume fraction of $%
x=0.14$ and $x=0.15$. (c) In-plane resistivity in zero field for Fe$_{1+y}$%
(Te$_{1-x}$S$_{x}$)$_{z}$ (c). Arrows show positions of peaks in $\partial 
\protect\rho /\partial T$ that correspond to anomalies in magnetization
associated with SDW transition (a,b).}
\end{figure}

\begin{figure}[tbh]
\centerline{\includegraphics[scale=0.3]{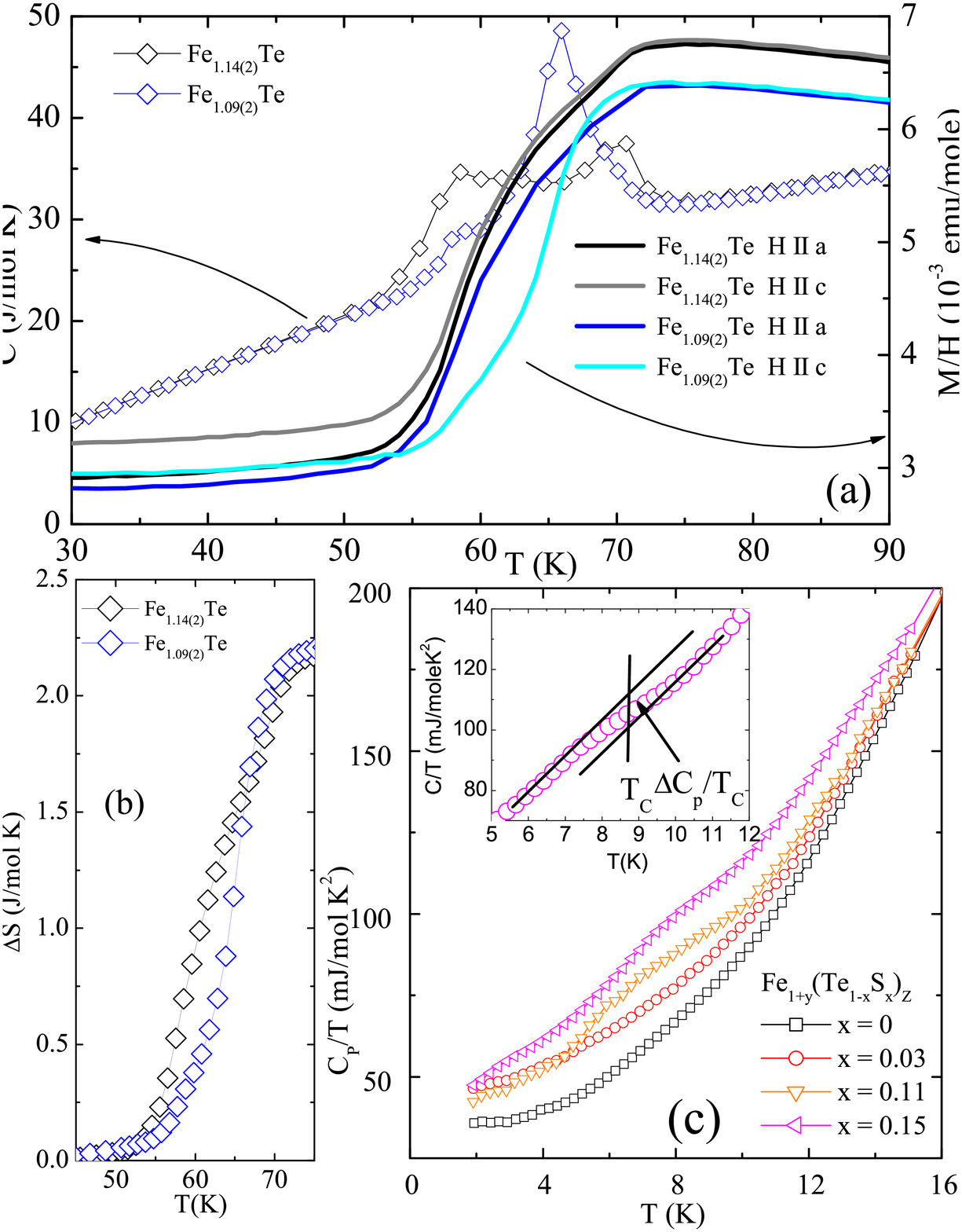}} 
\vspace*{-0.3cm}
\caption{(a) Magnetization and heat capacity C$_{p}$ at SDW transition of Fe$%
_{1.14(2)}$Te and Fe$_{1.09(2)}$Te. (b) Entropy balance around the
transition. (c) Low temperature heat capacity with discontinuity in C$_{p}$%
/T at T$_{C}$ for $x=0.15$ sample shown in the inset. Clear jumps associated
with superconducting transitions are seen for $x\geq 0.11$.}
\end{figure}

Fig. 2 (a,b) shows the anisotropic temperature dependence of magnetic
susceptibility for Fe$_{1+y}$(Te$_{1-x}$S$_{x}$)$_{z}$. The peak at 70 K for
Fe$_{1.14(2)}$Te corresponds to an antiferromagnetic (AF) transition
presumably coupled with structural and first order.\cite{Li} The transition
spans about 20 K for both field orientations. The magnetic susceptibility is
isotropic above 70 K and $\chi ^{\parallel c}/\chi ^{\perp c}$ increases
from 1 to 1.1 below the transition. The transition temperature is suppressed
with sulfur doping down to 20 K by $x=0.15$. A diamagnetic signal is
observed for $x\geq 0.14$ (Fig. 2(a,b) insets), in apparent coexistence with
a magnetic state. Magnetic susceptibility is Curie-Weiss like above 200 K.
The effective moments estimated are between the low spin ($2.94\mu _{B},$ $%
S=1$) and high spin ($4.9\mu _{B},$ $S=2$) values of an Fe$^{2+}$ (3d$^{6}$)
in a tetragonal crystal field (Table I). The high temperature effective
moments decrease with the S-doping and\ with the reduction of excess Fe.
Negative Curie-Weiss temperatures attest to the antiferromagnetic coupling
between moments (Table I).

The in-plane electrical resistivity in zero field is shown in Fig. 2(c).
Residual resistivity values at low temperatures for pure Fe$_{1+y}$Te and
crystals with the highest sulfur concentration $x$ are comparable to single
crystals grown by Bridgeman method,\cite{Sales} but smaller by a factor of 3
- 4 than in polycrystalline materials\cite{Mizuguchi} due to the absence of
grain boundaries and secondary phases. The grain boundaries are not
transparent as in MgB$_{2}$ where intrinsic low values of $\rho _{0}$ in
high quality polycrystals are often \ lower than in crystals.\cite{Raquel}
Therefore grain boundaries cannot be neglected when measuring resistivity on
polycrystals of iron chalcogenide superconductors. The resistivity of Fe$%
_{1+y}$Te above the magnetic transition is poorly metallic, in agreement
with measurements on polycrystals and an optical conductivity study which
did not find a semiconducting gap.\cite{Mizuguchi}$^{,}$\cite{Chen} This is
an important distinction from the metallic resistivity above $T_{c}$ in iron
based superconductors of ThCr$_{2}$Si$_{2}$, ZrCuSiAs structure or even FeSe.%
\cite{Hsu}$^{,}$\cite{Rotter}$^{-}$\cite{Wu} The magnitude of the
resistivity, $\rho $, becomes larger with small doping for $x$ = 0.03 but
decreases with the increase of S concentration. Two distinct contributions
to $\rho $ are observed in the low temperature phase at the temperature of
the magnetization anomaly: semiconducting and metallic or semimetallic
(arrows Fig 2(c)). For $x=0.03$ a small decrease in $\rho $ is observed
whereas the $x=0.1$ sample shows an increase in resistivity, indicating that
a part of the Fermi surface is destroyed. This can be understood within the
framework of density functional (DFT) calculations that predict a metal with
"nesting" cylindrical Fermi surfaces which are separated by a wavevector
corresponding to spin density wave (SDW) low temperature ground state.\cite%
{Subedi} With further increase of $x$, the resistivity anomaly at the
magnetic transition is smaller and broader when compared to $x=0$. Though
all samples for $x\geq 0.03$ show a clear onset of superconductivity; zero
resistivity is observed for $x\geq 0.10$ (Fig. 2(c)).

Figure 3(a) gives the temperature dependent specific heat C$_{p}$ and M/H
for Fe$_{1.14(2)}$Te and Fe$_{1.09(2)}$Te below 90 K. Both crystals show two
lambda anomalies at magnetic/structural transition around\ $T_{1}=70$ K and $%
T_{2}=59$ K for Fe$_{1.14(2)}$Te and around $T_{1}=66$ K and $T_{2}=59$ K
for Fe$_{1.09(2)}$Te. Above and below the transition region there is no
difference in C$_{p}$(T). A magnetic field of 90 kOe shifts both transition
in both samples for $\Delta T=1$ K. The entropy $\Delta S=2.2$ J/mol
associated with the transition is independent of the iron stoichiometry $y$
(Fig. 3(b)). This is smaller than estimated change of entropy in Fe$_{1.07}$%
Te of $\Delta S\sim 3.2$ J/mole(K).\cite{Li} The error is probably due to
conventional PPMS heat capacity setup which introduces sizeable error in the
vicinity of the first order phase transition.\cite{Lashley} Nevertheless, we
can still compare the change in $\Delta S$ for Fe$_{1+y}$Te crystals with
different $y$ caused by AF contribution which dominates $\Delta S$ in the
transition region.\cite{Li} Assuming that total entropy is lost on the spin
state transition $\Delta S=R\ln [(2S_{H}+1)/(2S_{L}+1)]$ and using $\mu
_{eff}=\sqrt{4S(S+1)}$, high temperature effective moment is $\mu
_{eff}^{H}( $Fe$_{1.14(2)}$Te$)=3.92$ $\mu _{B}$, $\mu _{eff}^{H}($Fe$%
_{1.09(2)}$Te$)=3.73$ $\mu _{B}$, we obtain the moment value below magnetic
transitions in the ordered state $\mu _{eff}^{H}($Fe$_{1.14(2)}$Te$)=1.3$ $%
\mu _{B}$ and $\mu _{eff}^{H}($Fe$_{1.09(2)}$Te$)=1.2$ $\mu _{B}$. Larger
relative entropy change for higher $y$ is related to the occupancy of iron
in the interstitial sites which is expected to be strongly magnetic.\cite%
{Zhang} Interestingly, these numbers are very close to values for a spin
moment of $1.3$ $\mu _{B}$ associated with SDW transition calculated by DFT
calculations.\cite{Subedi}

The low temperature C$_{p}$ data for Fe$_{1.14}$Te can be fitted to the $%
C(T)=\gamma T+\beta T^{3}$ power law below 15 K with $\gamma =32$ mJ/moleK$%
^{2}$ and $\beta =0.49$ mJ/moleK$^{4}$ from which a $\theta _{D}=228$ K can
be obtained (Fig. 3(c)). Specific heat shows a broad feature around $T_{c}$
(Fig. 3(c)) inset) for superconducting samples, similar to other iron
pnictides.\cite{Nini} Due to high upper critical fields and apparent
coexistence of superconductivity and long range magnetic order in our
crystals, a reliable estimate of the normal state contribution to electronic
specific heat $\gamma $ is rather difficult. Additional uncertainty in
testing traditional isotropic weak coupling BCS value of $\Delta
C_{p}/\gamma T_{c}$ in Fe$_{1+y}$(Te$_{1-x}$S$_{x}$)$_{z}$ is introduced by
the percolative nature of superconductivity with up to 7 \% superconducting
volume fraction (Fig. 2). Therefore we restrict ourselves to an estimate of
the C$_{p}$/T discontinuity associated with superconducting transition for
material with the highest sulfur concentration and consequently the most
pronounced jump in specific heat.\cite{Sergeyhc} For $x=0.15$ it is about 12
mJ/moleK$^{2}$ at $T_{c}$ = 8.8 K, comparable to what is observed in Ba(Fe$%
_{1-x}$Co$_{x}$)$_{2}$As$_{2}$ and Ba(Fe$_{1-x}$Ni$_{x}$)$_{2}$As$_{2}$
single crystals.\cite{Sergeyhc}

\begin{figure}[tbh]
\centerline{\includegraphics[scale=0.3]{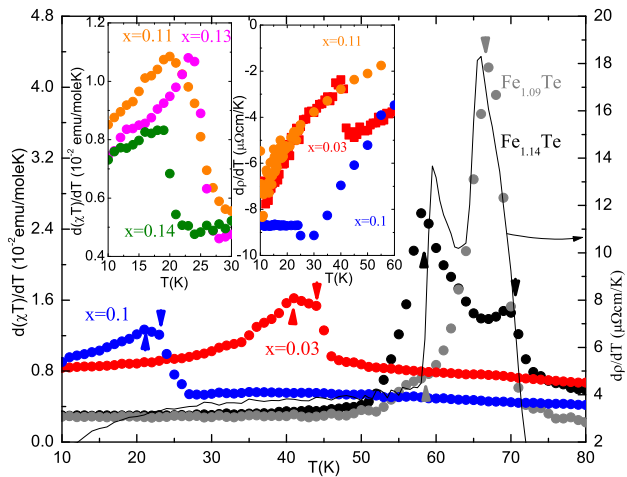}} 
\vspace*{-0.3cm}
\caption{Fe$_{1.14(2)}$Te and Fe$_{1.09(2)}$Te as well as sulfur doped
samples for $x\leq 0.1$ show two clear magnetization anomalies around SDW
transition, as seen in derivatives $d(\protect\chi T)/dT$ and $d\protect\rho %
/dT$ (inset). For higher sulfur content anomalies are broader (insets) and
cannot be distingushed.}
\end{figure}

Closer inspection of the ($\partial \chi T/\partial T$)\cite{Fisher} and $%
\partial \rho /\partial T$ data (Fig. 4) for Fe$_{1.14(2)}$Te and Fe$%
_{1.09(2)}$Te unveils two transitions at temperatures T$_{1}$ and T$_{2}$
that correspond to specific heat anomalies in Figure 3(a) (Table I). With
sulfur substitution both transitions are clearly observed only up to $x\leq
0.1$ (Fig. 4). For higher sulfur concentration only one broad anomaly can be
observed (Fig. 4 insets). Two successive transitions were reproduced on
independently grown crystals within the same batch and in different batches.
The exact temperatures of transitions T$_{1}$ and T$_{2}$ did vary from
batch to batch.

\section{Discussion}

The AF SDW in pure Fe$_{1+y}$Te is accompanied by a lattice distortion for
all investigated values of $y$ as in the undoped Fe-As superconductors.\cite%
{Rotter}$^{,}$\cite{Chen2}$^{-}$\cite{Klauss} DFT calculations have found
that excess Fe donates charge as Fe$^{+}$ to FeTe layers, with strong
tendency of moment formation on the excess Fe site.\cite{Zhang} By comparing
our C(T) data with the specific heat data taken on Fe$_{1.06}$Te crystals%
\cite{Sergeyte} it can be seen that the clarity of the two step anomaly
increases with the increase of $y$ in Fe$_{1+y}$Te. It is absent for Fe$%
_{1.05}$Te and Fe$_{1.06}$Te\cite{Chen2}$^{,}$\cite{Sergeyte}, visible for Fe%
$_{1.09}$Te and rather pronounced for Fe$_{1.14}$Te with similar entropy
under both transitions (Fig 3a,b). Magnetic measurements ($\partial \chi
T/\partial T$ )closely match thermodynamic data (Fig.3, Fig.4). Whereas
temperature of lower temperature transition T$_{2}$ (59 K) does not change
with change in $y$, Fe$_{1+y}$Te with higher content of excess Fe $y$ has
transition T$_{1}$ at higher temperature (70 K and 66 K, Table I). The Fermi
level in Fe$_{1+y}$Te lies exactly at the sharp peak of \ the excess Fe
density of states N(E$_{F}$); therefore higher T$_{1}$ may be magnetically
driven based on the Stoner criterion N(E$_{F}$)I > 1.\cite{Zhang} Increased $%
\rho $ values for Fe$_{1.14}$Te when compared to Fe$_{1.09}$Te are also
consistent with this (Fig. 2). Higher level of excess Fe $y$ corresponds to
larger size mismatch between cylindrical electron and hole Fermi surfaces.
Therefore T$_{2}$ and T$_{1}$ transitions may correspond to successive SDW
Fermi surface nesting of individual electron - hole cylindrical pieces.\cite%
{Subedi}$^{,}$\cite{Zhang} Recent work shows that the magnetic order in
parent compounds of iron based superconductors is established below
temperature of structural transition with up to 20 K difference in
temperature of transition, as seen in CeFe$_{1-x}$Co$_{x}$AsF .\cite{Cruz}$%
^{,}$\cite{XiaoY} It is unlikely that two transitions seen in our crystals
correspond to individual magnetic and structural transitions since they have
the same sensitivity to magnetic field.

Our findings are summarized in the electronic and structural phase diagrams
shown in Fig. 5. The lattice contraction with isoelectronic sulfur
substitution corresponds to a positive chemical pressure. The magnetic
transition is suppressed from the $\sim 58-70$ K region to about 20 K.
Signatures of percolative superconductivity were observed for all $x\geq 0.3$%
. Zero resistivity in fully percolating path was observed for $x\geq 0.1$.
The superconducting transition width decreases with the increase of $x$ and $%
T_{c}$. Clearly, there is a competition between magnetic SDW order and the
superconducting state since with increase in sulfur content $x$, $%
dT_{1,2}/dx $ and $dT_{c}/dx$ have opposite signs.

Having delineated the evolution of magnetic and superconducting properties,
it is natural to ask what is the correlation with the structural parameters.
The unit cell parameters \textit{a} and \textit{c} of \textit{P4/nmm}
crystal structure decrease smoothly at $T=80$ K as sulfur is substituted in
the place of tellurium (Table I). The \textit{c/a }ratio decreases to nearly
constant value for $x\geq 0.1$ up to $x=0.15$. After $x=0.15$ we have
observed formation of FeS in the hexagonal NiAs - type of structure in the
same range of synthesis parameters. Close inspection of the tetrahedral
angle $\alpha $ at $T=80$ K (Fig. 5(b)) reveals an extremum near the
superconducting percolation threshold. The angle $\alpha $ increases up to $%
x=0.03$ and then decreases with further sulfur increase. The tetrahedral
angle $\alpha $ therefore seems to be intimately connected with electronic
transport properties which will be discussed next.

Both $x=0$ crystals are metallic in the low temperature phase (Fig. 2(c)).
On the other hand, two successive transitions have also been reported in FeTe%
$_{0.92}$ under high pressure in the intermediate regime between $P=(1-1.8)$
GPa,\cite{Okada} as well as two distinct types of transport below the
magnetic and structural transition:\ metallic for FeTe$_{0.9}$ and
semiconducting for FeTe$_{0.82}$.\cite{Fang} We note that semiconducting
contribution to $\rho $ below the magnetization anomaly for $x=0.10$ and $%
x=0.11$ (Fig. 1) coincides with Te(S) vacancies from synchrotron X ray
refinement (Fig. 1, Table I). Crystals with no Te(S) vacancies within error
bars have metallic or semimetallic contributions to $\rho $. This is in
agreement with photoemission studies that showed no visible energy gaps at
the electron and hole Fermi surface for $y<0.05$ in Fe$_{1+y}$Te.\cite{Xia}
Increase of resistivity at the SDW AF transition signals small gap opening
at the Fermi surface. The band structure of FeTe features intersecting
elliptical cylindrical electron portions at the Brillouin zone corners
compensated by hole sections with higher effective mass at the zone center.%
\cite{Subedi} Our findings show that the details of the nesting condition
depend rather sensitively on the tetrahedral angle $\alpha $ and vacancies
on the ligand site. This points to importance of hybridization between Te $p$
and Fe $d$ bands in addition to excess stoichiometry $y$ on Fe site.\cite%
{Han} Our results strongly suggest that nanoscale inhomogeneity seems to be
the key factor governing magnetic and electronic transport properties in Fe$%
_{1+y}$(Te$_{1-x}$S$_{x}$)$_{z}$.

\begin{figure}[tbh]
\centerline{\includegraphics[scale=0.3]{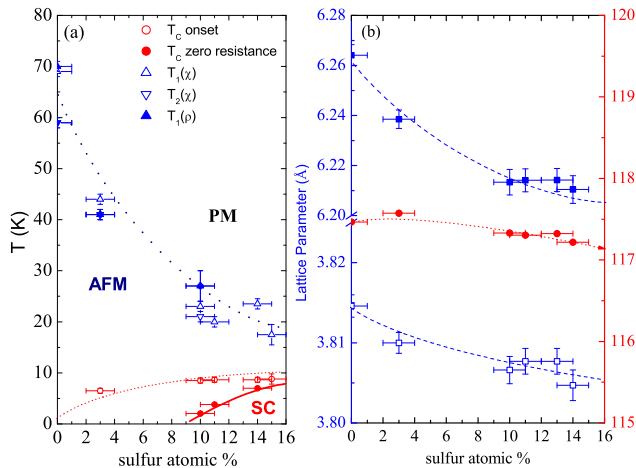}} 
\vspace*{-0.3cm}
\caption{(a) Electronic phase diagram of Fe$_{1+y}$(Te$_{1-x}$S$_{x}$)$_{z}$%
, showing paramagnetic (PM), antiferromagnetic (AFM) and superconducting
(SC) ground states. Blue triangles pointing up and down correspond to T$_{2}$
and T$_{1}$ transitions respectively. For $x=0$ both can easily be
identified. Red circles denote onset of superconducting transition in $%
\protect\rho $ and zero resistance. Transition for $x=0.15$ was estimated
from heat capacity measurement. (b) Structural parameters at $T=80K$.}
\end{figure}

Finally we comment on the percolative nature of superconductivity found in
our crystals. Superconducting volume fraction increases with sulfur
stoichiometry $x$. The $\ 4\pi \chi _{v}$ reaches up to $-0.07$ at T=1.8 K ($%
\sim $ 0.26 $T_{c}$) for the highest $x$ crystals where zero resistivity was
observed to approach the $T_{c}$ onset (Fig. 5) . This signals granular
superconducting state coexisting with SDW order, taking only a fraction of
sample volume and stabilizing to fully percolating superconducting path by $%
x=0.14$. Similar coexistence was observed in other iron based
superconductors, CaFe$_{1-x}$Co$_{x}$AsF, SmFeAsO$_{1-x}$F$_{x}$, SrFe$_{2}$%
As$_{2}$ and BaFe$_{2}$As$_{2}$.\cite{Drew}$^{,}$\cite{XiaoY}$^{,}$\cite%
{Colombier} For example in the underdoped region of Ba$_{1-x}$K$_{x}$Fe$_{2}$%
As$_{2}$ the superconducting volume fraction has been reported to be (23$\pm 
$3)\% of -1/4$\pi $ at $\sim $ 0.06$T_{c}$ increasing up to 50\% for nearly
optimally doped material.\cite{Nini}$^{,}$\cite{Goko}$^{,}$\cite{Park} Since
SDW magnetic order and superconductivity compete for the same Fermi surface,
percolative nature of superconductivity may be associated with intrinsic
mesoscopic real space phase separation as in cuprate oxides or CaFe$_{1-x}$Co%
$_{x}$AsF.\cite{XiaoY}$^{,}$\cite{Jorgensen}$^{,}$\cite{Pan} Consequently
superconductivity may be mediated by magnetic fluctuations, consistent with
small values of electron phonon coupling constant found in doped Fe$_{1+x}$%
Te and FeSe.\cite{Subedi}

\section{Conclusion}

In conclusion, we have performed combined and comprehensive study of
structural, magnetic and superconducting properties of Fe$_{1+y}$(Te$_{1-x}$S%
$_{x}$)$_{z}$ single crystals. Magnetic transition decreases from $\sim $
58-70 K\ to about 20 K for $x=0.15$. We have shown that the increase of
excess Fe $y$ in Fe$_{1+y}$Te results in two anomalies in thermodynamic,
magnetization and transport properties. Electronic transport is rather
sensitive to possible vacancies on Te(S) site. Filamentary superconductivity
is observed for all $x$ in apparent coexistence with magnetism. Microscopic
measurements such as muon spin rotation ($\mu $SR) are needed to confirm
real space phase separation and/or coexistence of superconductivity and
magnetism. High pressure synthesis techniques could increase the
superconducting volume fraction. That would enable the answer to the
question if maximal T$_{C}$ occurs at the point where magnetism disappears
as in cuprate oxides and other complex iron arsenide superconductors.

\section{Acknowledgments}

We are grateful for helpful discussions with Sergey Bud'ko, Simon Billinge
and Myron Strongin. This work was carried out at the Brookhaven National
Laboratory, which is operated for the U.S. Department of Energy by
Brookhaven Science Associates (DE-Ac02-98CH10886). Use of the Advanced
Photon Source was supported by the U. S. Department of Energy, Office of
Science, Office of Basic Energy Sciences, under Contract No.
DE-AC02-06CH11357. This work was supported by the Office of Basic Energy
Sciences of the U.S. Department of Energy.

\end{document}